\documentclass[pra,showkeys,amssymb,amsmath,twocolumn]{revtex4}

\begin{document}

\title{Concurrence-based entanglement measure for Werner States}
\author{Kai Chen$^{1}$}
\author{Sergio Albeverio$^{1}$}
\author{Shao-Ming Fei$^{1,2}$}
\affiliation{$^1$Institut f\"ur Angewandte Mathematik, Universit\"at Bonn, D-53115,
Germany\\
$^2$Department of Mathematics, Capital Normal University, Beijing 100037,
China}

\begin{abstract}
We give explicit expressions for entanglement measures of Werner
states in arbitrary dimensions in terms of concurrence and tangle.
We show that an optimal ensemble decomposition for a joint density
matrix of a Werner state can achieve the minimum average
concurrence and tangle simultaneously. Furthermore, the same
decomposition also attains entanglement of formation for Werner
states.
\end{abstract}


\maketitle

Quantum entanglement is playing very significant roles in quantum information
processing such as quantum cryptography, quantum teleportation and quantum
computation \cite{nielsen}. This motivates an increasing interest in the study
of operational detection and quantification of entanglement for various quantum systems.
Despite of a great deal of efforts in recent years,
for the moment only partial solutions are known to detect and quantify
entanglement for generic mixed state.

The crucial entanglement measure \emph{concurrence}, firstly proposed by Hill and
Wootters \cite{Hill-Wootters97,Wootters98},
has recently been shown to play
an essential role in describing quantum phase transitions in various
interacting quantum many-body systems \cite{Osterloh02-Wu04}, affecting
macroscopic properties of solids significantly \cite{Ghosh2003} and revealing
distinct scaling behavior for different types of multipartite entanglement
\cite{Carvalho05}. The concurrence was then generalized
by Uhlmann, Rungta \textsl{%
et al}, and by Albeverio and Fei \cite{Rungta01-AlbeverioFei01} to arbitrary
bipartite quantum system. Multi-variable concurrence vectors are also
introduced in \cite{Audenaert01,Multi-variables} and possible multipartite
generalizations are given in \cite{multigen}.

However, even the problem of obtaining only lower bound of concurrence has
required considerable efforts \cite{lowerbound,Audenaert01}. This
problem has been advanced significantly in
\cite{Mintert04-MintertPhD}, providing an algebraic lower bound
which can be optimized further by numerical approaches, and in
\cite{Chen-Albeverio-Fei-PRL200504} through an entirely analytical
derivation of a complementary tightly lower bound. In addition,
nice analytical results are also given for isotropic states
\cite{Rungta-Caves03} and rotationally symmetric states
\cite{Manne-Caves2005}.

An important class of quantum states are the Werner states \cite%
{werner89,Vollbrecht-Werner01}, which appear in realistic quantum computing
devices and quantum communication environments, e.g.
transmitting perfect entangled states through a noisy depolarizing channel.
An effective experimental generation of these states has been recently demonstrated
in \cite{wernerpreparation}. An analytical expression has been derived in
\cite{Vollbrecht-Werner01} for entanglement of formation (EOF), which
quantifies the minimally required physical resources to prepare a Werner state.
The greatest cross norm is also obtained for the Werner states \cite{Rudolph02}.
It is believed \cite{WoottersQIC2001} that there is a novel connection between
the concurrence and their EOF, through a parameter that depicts the Werner state
completely. One expects that the situation would be similar to the case of two qubits
where EOF is an analytic monotone function of concurrence \cite{Wootters98}.
However, for Werner states why such a parameter plays the role
of concurrence is not yet well understood. There is also no rigorous and clear
proof of this fact in the literatures, for high dimensions.

In this letter we first find an analytic
expression of concurrence for Werner states in {\em arbitrary dimensions}, which
complements many of the existing analytic results. We then show how
EOF is exactly related to the concurrence. We demonstrate that,
surprisingly, an optimal ensemble decomposition will achieve concurrence,
tangle and EOF simultaneously for the Werner states. Thus the expected
connection is rigorously proved and shown to be natural.

\vspace{0.2cm}
\textbf{Werner states.}
The Werner states are a class of mixed states for $d\times d$ systems (two
qudits with $d\geq 2$) which are invariant under the transformations ${U}{\otimes U}$,
for any unitary transformation $U$ \cite{werner89,Vollbrecht-Werner01}. The
density matrix of these states can be expressed as
\begin{equation}
\rho _{f}={\frac{{1}}{d^{3}-d}}(d-f)\mathbb{I}+(df-1)\mathbb{F},
\label{wernerstate}
\end{equation}
where $\mathbb{F}$ is the flip operator (or swap operator) defined by
$\mathbb{F}(\phi \otimes \psi )=\psi \otimes \phi $. In the computational
basis $|ij\rangle $, $\mathbb{F}$ is of the form
$\mathbb{F}=\sum_{i,j}^{d}|ij\rangle \langle ji|$. Here
$f$ is a constant  $f=\langle \mathbb{F}
\rangle \equiv Tr(\mathbb{F}\rho _{f})$  satisfying $-1\leq f\leq 1$.
Werner states are separable if and only if $f\geq 0$, as shown
in \cite{werner89,Vollbrecht-Werner01}.

It is evident that the Werner states are invariant under the action of the
LOCC ``twirling" superoperator $\mathcal{T}$
\begin{equation}
\mathcal{T}(\rho _{f})=\int dU\,U\otimes U\rho _{f}U^{\dagger }\otimes {U}%
^{\dagger }=\rho _{f}\;.  \label{twirling}
\end{equation}
where $dU$ refers to the standard Haar measure on the unitary
matrix group. Consider an arbitrary initial pure $d\otimes d$ quantum state
of the standard Schmidt form
\begin{equation}
{\left\vert \psi \right\rangle }=\sum_{ij}\Phi _{ij}{\left\vert
ij\right\rangle =}\sum_{i}\sqrt{\mu _{i}}{\left\vert a_{i}b_{i}\right\rangle
=}\big(U_{A}\otimes U_{B}\big)\sum_{i}\sqrt{\mu _{i}}{\left\vert
ii\right\rangle },  \label{Schmidt}
\end{equation}
where ${\left\vert a_{i}\right\rangle }$ and ${\left\vert b_{i}\right\rangle }$ are
orthonormal bases of the subsystems $\mathcal{H}_{A}$ and $\mathcal{H}_{B}$, respectively.
The state ${\left\vert \psi \right\rangle }$ is thus specified
by its Schmidt vector $\vec{\mu}%
\equiv (\mu _{1},\mu _{2},\ldots ,\mu _{d})$ and the unitary operators $%
U_{A}$ and $U_{B}$. For convenience of later use, we use the symbol $%
\Phi $ to denote the pure state ${\left\vert \psi \right\rangle }$, where
$\Phi $ is the matrix with entries $\Phi_{ij}$, which contains all the information for ${%
\left\vert \psi \right\rangle }$. In fact, any two-qudit $\rho $ reduces to
a Werner state
\begin{equation}
\mathcal{T}(\rho )=\rho _{f(\rho )},
\end{equation}
under the twirling superoperator while
keeping $f(\rho )=\langle \mathbb{F}%
\rangle =Tr(\mathbb{F}\rho )$ invariant. This can easily be seen from
\begin{equation}
\text{Tr}\big(\mathbb{F}\mathcal{T}(\rho )\big)=\text{Tr}\big(\mathcal{T}%
\mathbb{(F)}\rho \big)=\text{Tr}(\mathbb{F}\rho ).
\end{equation}
As for the pure state Eq.~(\ref{Schmidt}), one has
\begin{equation}
\mathcal{T}(|{\psi }\rangle \langle {\psi }|)=\rho _{f},  \label{twirlpure}
\end{equation}
with $f$ given by
\begin{eqnarray}
f&=&Tr(|{\psi }\rangle \langle {\psi }|\mathbb{F})=\sum_{ij}\Phi _{ij}\Phi
_{ji}^{\ast }. \label{FEV}
\end{eqnarray}

\vspace{0.2cm}
\textbf{Entanglement measures in terms of concurrence and tangle.}
The (generalized) definition \cite{Rungta01-AlbeverioFei01} of
concurrence $C({\left\vert \psi \right\rangle })$ for a pure
state ${\left\vert \psi \right\rangle }$
is as follows: $C({\left\vert \psi \right\rangle })=%
\sqrt{2(1-\mbox{Tr}\rho _{A}^{2})}$, where the reduced density matrix $\rho
_{A}$ is given by $\rho ^{A}=tr_{B}({\left\vert \psi \right\rangle }{%
\left\langle \psi \right\vert })$. This
can then be extended to mixed states by the ``convex roof construction",
\begin{equation}
C(\rho )\equiv \min_{\{p_{i},|\psi _{i}\rangle \}}\sum_{i}p_{i}C({\left\vert
\psi _{i}\right\rangle }),  \label{defconcurrence}
\end{equation}
where $\rho =\sum_{i}p_{i}|\psi _{i}\rangle
\langle \psi _{i}|$, $p_{i}\geq 0$ and $\sum_{i}p_{i}=1$. For any pure
product state ${\left\vert \psi \right\rangle }$, $C({\left\vert \psi
\right\rangle })$ vanishes according to the definition. Consequently, a
state $\rho $ is \emph{separable} if and only if $C(\rho )=0$.
A separable state can then be
represented as a convex combination of product states \cite{werner89}.

Another entanglement measure called \emph{tangle}, was first proposed in
\cite{Coffman2000}. Its generalization to generic mixed states and
further properties were explored in \cite{Osborne2002,Rungta-Caves03}.
The tangle $\tau (\rho )$ is by definition the squared concurrence for pure states, and
can be similarly extended to mixed states
\begin{equation}
\tau (\rho )\equiv \min_{\{p_{i},|\psi _{i}\rangle \}}\sum_{i}p_{i}C^{2}({%
\left\vert \psi _{i}\right\rangle }),  \label{deftangle}
\end{equation}
where $C^{2}({\left\vert \psi _{i}\right\rangle })$ stands for
$\big(C({\left\vert \psi_{i}\right\rangle })\big)^{2}$.

For the pure state ${\left\vert \psi \right\rangle }$ of (\ref{Schmidt}), we have:
\begin{eqnarray}
\tau (\Phi ) &=&C^{2}(\Phi )=2\big(1-\sum_{i}\mu _{i}^{2}\big)  \nonumber \\
&=&4\sum_{i<j}\mu _{i}\mu _{j}=C^{2}(\vec{\mu})=\tau (\vec{\mu}),
\label{representations}
\end{eqnarray}
which varies smoothly from $0$, for pure product states, to $2(d-1)/d$ for
maximally entangled pure states.

\vspace{0.2cm}
\textbf{Concurrence and tangle for Werner states.}
To derive the tangle and concurrence for Werner states,
we will use a technique developed in
\cite{Vollbrecht-Werner01,Terhal-Voll2000,Rungta-Caves03}.
The EOF is defined to be $E(\rho )\equiv \min_{\{p_{i},|\psi _{i}\rangle
\}}\sum_{i}p_{i}E({\left\vert \psi _{i}\right\rangle })$ for all
possible ensemble realizations $\rho =\sum_{i}p_{i}|\psi
_{i}\rangle \langle \psi _{i}|$, where $p_{i}\geq 0$ and
$\sum_{i}p_{i}=1$. Here $E(\left\vert \psi \right\rangle )=S(\rho
_{A})$ with $S(\rho _{A})$  the entropy
${S(\rho _{A})}\equiv -\sum_{i=1}^{d}\mu _{i}\log _{2}\mu _{i}=H(\vec{\mu})$,
where $\mu _{i}$ are all the eigenvalues of ${\rho _{A}}$ and $\vec{\mu}$
is the Schmidt vector $(\mu _{1},\mu _{2},\ldots ,\mu _{d})$.
The EOF of Werner states is derived in \cite{Vollbrecht-Werner01} as being given by
\begin{equation}
E(\rho _{f})=H_{2}\big(\frac{1}{2}(1-\sqrt{1-f^{2}})\big),  \label{WernerEOF}
\end{equation}
by an elegant extremization procedure. Here $H_{2}(.)$ is the binary entropy
function. Since $E(\rho _{f})$ is a monotonically increasing function of $-f$, as seen from
Eq.~(\ref{WernerEOF}), it is expected \cite{WoottersQIC2001} that $-f$ plays
the role of concurrence, similarly as in the two qubits case \cite{Wootters98}.

\vspace{0.2cm}
\textsl{Simplification through symmetry.}
Before getting through possible extremization procedures, we
first recall some formulations of the convex roof construction of
entanglement measures \cite{Vollbrecht-Werner01,Soojoon2003}. We
denote by $K$ the whole set of states in a given quantum system
and by $M$ the set of all pure states in $K$.
Then the elements of K are convex linear combinations of
a finite number elements of $M$. Let $G$ be a compact group of
symmetries acting on $K$ by transformations $\alpha _{U}:\rho
\mapsto U\rho U^{\dagger } $, $U$ being an element of $G$, and
assume that a pure-state measure $E$ defined on $M$ is invariant
under $G$. We define a projection
$\mathbf{P}:K\rightarrow K$ by $\mathbf{P}\rho =\int dUU\rho
U^{\dagger }$ with $dU$, as before, the standard Haar measure on
$G$, and a function $\varepsilon $ on $\mathbf{P}K$ by
\begin{equation}
\varepsilon (\rho )=\min \{E({\left\vert \psi \right\rangle }):{\left\vert
\psi \right\rangle }\in M,\mathbf{P}{\left\vert \psi \right\rangle }{%
\left\langle \psi \right\vert }=\rho \}.  \label{measure}
\end{equation}
For $\rho \in \mathbf{P}K$, it is proved that \cite{Vollbrecht-Werner01}
\begin{equation}
\mathrm{co}E(\rho )=\mathrm{co}\varepsilon (\rho ),  \label{cohull}
\end{equation}
where $\mathrm{co}f$ at the right hand side stands for the convex hull construction
for a given function $f$ restricted to the pure states satisfying
$\mathbf{P}{\left\vert \psi \right\rangle }{\left\langle \psi%
\right\vert }=\rho$, as shown in Eq.~(\ref{measure}).
When we take the concurrence as the entanglement measure, $\mathrm{co}E(\rho )$ at
the left hand side of Eq.~(\ref{cohull}) corresponds to
\begin{eqnarray}
C(\rho ) &=&\mathrm{co}C(\rho )  \nonumber  \label{convexhull} \\
&=&\min \Biggl\{\sum_{i}p_{i}C({\left\vert \psi _{i}\right\rangle })\Biggm|%
\rho =\sum_{i}p_{i}|\psi _{i}\rangle \langle \psi _{i}|\Biggr\},  \nonumber
\\
&&
\end{eqnarray}
where the infimum is taken over all possible convex combinations with $p_{i}\geq 0$ and $%
\sum_{i}p_{i}=1$.

According to the above results, in order to derive concurrence or
tangle for the Werner states one thus needs only to consider all
the pure states $\sigma $ satisfying $P\sigma =\rho $ and
achieving minimal admissible concurrence or tangle for $\sigma $.
Finally one computes their convex hull. Here and later by
``minimal admissible", we mean the minimal value of concurrence or
tangle among all possible ensemble decompositions of the density
matrix.

\vspace{0.2cm}
\textsl{Extremization for pure states.}
With a given $f$ and the corresponding Werner
state $\rho _{f}$, we are going to find the desired pure states $\sigma $ with
coefficient matrix $\Phi$ satisfying $P\sigma =\rho _{f}$ and
minimize $C^{2}(\Phi )$. The task amounts to the following problem:
\begin{equation}
\left\{
\begin{array}{ll}
\text{minimize} & C^{2}(\Phi ) \\[1mm]
\text{subject to} & \sum_{ij}\Phi _{ij}\Phi _{ji}^{\ast }=f, \\[1mm]
& \sum_{ij}\left\vert \Phi _{ij}\right\vert ^{2}=1.
\end{array}
\right.   \label{optiprob1}
\end{equation}
The key point of our idea is to apply the concavity
properties of both $\tau (\Phi )$ and $C(\Phi )$ with respect to the reduced
density matrix $\rho ^{A}=tr_{B}({\left\vert \psi \right\rangle }{%
\left\langle \psi \right\vert })$, as proved in \cite{Rungta-Caves03}, i.e.
\begin{equation}
g(\lambda _{1}\Phi _{1}+\lambda _{2}\Phi _{2})\geq \lambda _{1}g(\Phi
_{1})+\lambda _{2}g(\Phi _{2}),  \label{concavity}
\end{equation}
where $\lambda _{1},\lambda _{2}\geq 0,\lambda _{1}+\lambda _{2}=1$ and where $g$
can be $\tau (\Phi )$ as well as $C(\Phi )$. By using this property, we will
derive tight lower bounds for $\tau (\Phi )$ resp. $C(\Phi )$, and then find a
condition under which the bound is achieved. Thus an essential step for
the minimization problem Eq.~(\ref{optiprob1}) is to find such a condition
under which the tight lower bound is achieved.

From Eq.~(\ref{Schmidt}), one has the reduced density
matrix $\rho ^{A}=\Phi \Phi ^{\dag }$. The $U\otimes U$ transformations
will neither change the degree of entanglement of a state nor the constraint condition
$\sum_{ij}\Phi _{ij}\Phi _{ji}^{\ast }=f$. In fact, it corresponds to a local
unitary transformation in $\rho ^{A}$, i.e., $\rho ^{A}\longrightarrow U\rho
^{A}U^{\dag }$. Thus one can choose conveniently $\Phi $ such as to make
$\rho ^{A}$ diagonal. The eigenvalues of $\rho ^{A}$ are then $%
\mu _{i}=\rho _{ii}^{A}=\sum_{k}\left\vert \Phi _{ik}\right\vert ^{2}$.
From Eq.~(\ref{representations}), the tangle is of the form
\begin{equation}
\tau (\Phi )=2\Big(1-\sum_{i}\big(\sum_{k}|\Phi _{ik}|^{2}\big)^{2}\Big).
\label{tangleeq}
\end{equation}

It is helpful to look at the eigenvalues of $\rho ^{A}$ as a distribution
of $d$ random variables
\begin{equation}
S=\Big(\sum_{k}\left\vert \Phi _{1k}\right\vert ^{2},\sum_{k}\left\vert \Phi
_{2k}\right\vert ^{2},\ldots ,\sum_{k}\left\vert \Phi _{dk}\right\vert ^{2}%
\Big),  \label{distri}
\end{equation}
which is a convex combination of the distributions
\begin{eqnarray}
&&S_{ij}=(\underset{i-1}{\underbrace{0,\ldots ,0}},\left\vert \Phi
_{ij}\right\vert ^{2},\underset{j-i-1}{\underbrace{0,\ldots ,0}},\left\vert
\Phi _{ji}\right\vert ^{2},\underset{d-j}{\underbrace{0,\ldots ,0}})/p_{ij},
\nonumber \\
&&\text{ \ \ \ \ \ \ \ \ \ \ \ with probability }p_{ij}=\left\vert \Phi
_{ij}\right\vert ^{2}+\left\vert \Phi _{ji}\right\vert ^{2},  \label{distri1}
\\
&&S_{ii}=(\underset{i-1}{\underbrace{0,\ldots ,0}},1,\underset{d-i}{%
\underbrace{0,\ldots ,0}}),  \nonumber \\
&&\text{ \ \ \ \ \ \ \ \ \ \ \ with probability }p_{ii}=\left\vert \Phi
_{ii}\right\vert ^{2},  \label{distri2}
\end{eqnarray}
where $\sum_{i\leq j}p_{ij}=1$ and $i<j\leq d$. Hence
\begin{equation}
S=\sum_{i\leq j}p_{ij}S_{ij}.  \label{spartition}
\end{equation}
Exploiting the concavity property of $\tau(\Phi )$, we get
\begin{equation}
\tau (S)\geq \sum_{i\leq j}p_{ij}\tau (S_{ij})=\sum_{i<j}p_{ij}\tau (S_{ij}),
\label{tangleconcavity}
\end{equation}
where we have used $\tau (S_{ii})=0$. On the other hand,
the function $f$ in Eq.~(\ref{FEV}) can be similarly expressed as
\begin{equation}
f=\sum_{i\leq j}p_{ij}f_{ij},  \label{fpartition}
\end{equation}
where
\begin{eqnarray}
f_{ij} &=&(\Phi _{ij}\Phi _{ji}^{\ast }+\Phi _{ji}\Phi _{ij}^{\ast })/p_{ij}
\nonumber \\
&=&2\text{Re}(\Phi _{ij}\Phi _{ji}^{\ast })/p_{ij},\text{ \ \ \ \ \ \ \ for}%
\ i<j  \label{fij} \\
f_{ii} &=&1.  \label{fii}
\end{eqnarray}
We now look for a lower bound of $%
\tau (S_{ij})$ for a given $f_{ij}$. Set $x=\Phi _{ij}/%
\sqrt{p_{ij}}$, $y=\Phi _{ji}/\sqrt{p_{ij}}$, Minimizing $\tau (S_{ij})$ is
equivalent to
\begin{equation}
\left\{
\begin{array}{ll}
\text{minimize} & \tau (S_{ij})=4|xy|^{2}=4|x|^{2}(1-|x|^{2}) \\[1mm]
\text{subject to} & 2\text{Re}(xy^{\ast })=f_{ij}, \\[1mm]
& |x|^{2}+|y|^{2}=1.
\end{array}
\right.   \label{subminproblem}
\end{equation}
Since $4|x|^{2}(1-|x|^{2})$ is a monotonically increasing function of $%
|x|^{2}$ taking values from $0$ to $1/2$, minimizing $\tau (S_{ij})$ is
equivalent to minimizing $|x|^{2}$
for given $f_{ij}$. This kind of problem was solved in \cite{Vollbrecht-Werner01},
and the solution is $|x|_{\min }^{2}=$ $(1-\sqrt{1-f_{ij}^{2}})/2$. Thus
$\tau (S_{ij})\geq 4|x|_{\min }^{2}(1-|x|_{\min }^{2})=f_{ij}^{2}$. From
Eq.~(\ref{tangleconcavity}), one has further
\begin{eqnarray}
\tau (S) &\geq &\sum_{i<j}p_{ij}\tau (S_{ij})\geq
\sum_{i<j}p_{ij}f_{ij}^{2}\geq \Big(\sum_{i<j}p_{ij}f_{ij}\Big)^{2}
\nonumber \\
&=&\big(f-\sum_{i}p_{ii}\big)^{2}=\Big(f-\sum_{i}|\Phi _{ii}|^{2}\Big)^{2},
\label{lbound}
\end{eqnarray}
where we have used the convexity property of $f_{ij}^{2}$ in the third
inequality of Eq.~(\ref{lbound}).

\vspace{0.2cm} \noindent \emph{Case 1: $f\geq 0$}

$\tau (S)$ itself vanishes if there is {\em only one} nonzero eigenvalue $1$
of $\rho ^{A}$, say
\begin{equation}
\rho _{ii}^{A}=\sum_{k}\left\vert \Phi _{ik}\right\vert ^{2}=1.
\label{cond1}
\end{equation}
The minimum $(f-\sum_{i}\left\vert \Phi _{ii}\right\vert ^{2})^{2}$ will
be $0$ if one chooses in addition
\begin{equation}
\left\vert \Phi _{ii}\right\vert ^{2}=f.  \label{cond2}
\end{equation}
The two equations Eqs.~(\ref{cond1}) and (\ref{cond2}) can always be
satisfied by a suitable choice of $\Phi $. Thus the minimal admissible
value for $\tau (S)$ is $0$.

\vspace{0.2cm} \noindent \emph{Case 2: $f<0$}

It is clear that any choice of nonzero $\Phi _{ii}$ will increase the value
of $(f-\sum_{i}\left\vert \Phi _{ii}\right\vert ^{2})^{2}$. Therefore for an
optimal solution one should have all $\Phi _{ii}=0$, if possible. On the other
hand, the equalities in Eq.~(\ref{lbound}) hold, if there is one
single item in the summation, due to the concavity property of $\tau (S)$
(the first inequality in Eq.~(\ref{lbound})) and
the convexity of $f^{2}$ (the third inequality in Eq.~(\ref{lbound})).
This is because all the inequalities will become equalities
$\tau (S)=\tau (S_{ij})=f_{ij}^{2}=f^{2}$ when
$p_{ij}=1$. Therefore we have two
nonzero components left, say $\Phi _{ij}$ and $\Phi _{ji}$. Hence one has
\begin{eqnarray*}
\Phi _{ij}\Phi _{ji}^{\ast }+\Phi _{ji}\Phi _{ij}^{\ast } &=&2\text{Re}(\Phi
_{ij}\Phi _{ji}^{\ast })=f, \\
\left\vert \Phi _{ij}\right\vert ^{2}+\left\vert \Phi _{ji}\right\vert ^{2}
&=&1,
\end{eqnarray*}
and thus
\begin{equation}
\left.
\begin{array}{c}
\Phi _{ij}=e^{i\theta _{1}}\big((1-\sqrt{1-f^{2}})/2\big)^{\frac{1}{2}}, \\
\Phi _{ji}=e^{i\theta _{2}}\big((1+\sqrt{1-f^{2}})/2\big)^{\frac{1}{2}},%
\end{array}
\right.   \label{solution}
\end{equation}
where $\theta _{1,2}$ are arbitrary real numbers satisfying
$\theta _{1}-\theta _{2}=(2n+1)\pi $, with $n$ being any integer.
With these choices of $%
\Phi _{ij}$ and $\Phi _{ji}$ in Eq.~(\ref{solution}), one gets the
minimal admissible value of $\tau _{\min }(S)=f^{2}$.

Combining all the above results, we have
\begin{equation}
\tau _{\min }(\Phi )=\left\{
\begin{tabular}{lll}
$f^{2},$ & \ \  & $\text{for }f<0$ \\
$0,$ & \ \ \  & $\text{for }f\geq 0$
\end{tabular}
\right.   \label{tmin}
\end{equation}
for $\Phi $ satisfying $P{\left\vert \Phi \right\rangle }{\left\langle \Phi
\right\vert }=\rho _{f}$.

\vspace{0.2cm} Since $C(\Phi )$ is also concave and is a monotonously
increasing function of $\tau (\Phi )$, we have
similar expressions as in Eqs.~(\ref{tangleconcavity}) and (\ref{lbound}),
\begin{eqnarray}
C(S) &\geq &\sum_{i\leq j}p_{ij}C(S_{ij})=\sum_{i<j}p_{ij}C(S_{ij})
\nonumber \\
&\geq &\sum_{i<j}p_{ij}|f_{ij}|\geq \biggl|\sum_{i<j}p_{ij}f_{ij}\biggr|
\nonumber \\
&=&\biggl|f-\sum_{i}p_{ii}\biggr|=\biggl|f-\sum_{i}|\Phi _{ii}|^{2}\biggr|.
\label{concurrencemin}
\end{eqnarray}
The above analysis for minimizing tangle $\tau (S)$ can naturally be
extended to the minimization of $C(S)$. It is evident that the solution of
Eq.~(\ref{subminproblem}) also achieves the minimal admissible value
\begin{equation}
C_{\min }(\Phi )=\left\{
\begin{array}{ll}
\left\vert f\right\vert =-f, & \text{for\ }f<0 \\[1mm]
0, & \text{for\ }f\geq 0%
\end{array}%
\right.   \label{cmin}
\end{equation}
for $\Phi $ satisfying $P{\left\vert \Phi \right\rangle }{\left\langle \Phi
\right\vert }=\rho _{f}$.

\vspace{0.2cm} \noindent \emph{Remark: }It is shown in \cite%
{Virmani-PlenioPLA2000} that different entanglement measures will
produce the same ordering for pure states if they reduce to the
entropy of entanglement for pure states. However, the concurrence
and tangle do not belong to that class. In fact, they will
generally lead to different orderings when compared with EOF for
pure states, since there are no simple monotonous function
relations among them and the EOF $E(\Phi )$ (except
for an apparent connection $E(\Phi )=H_{2}\big(\frac{1}{2}(1-\sqrt{%
1-C(\Phi )^{2}})\big)$ holding only for $2\otimes N$ systems as
easily seen from the definition). This means that a state $\Phi $
achieving a minimal $E(\Phi )$ may not automatically produce a minimal
$\tau (\Phi )$ or $C(\Phi )$. In our
case of Werner states, it occurs by chance that the solution Eq.~(\ref%
{solution}) achieves minima for all of the three entanglement
measures.

\vspace*{10pt}
With the above derived results, we can now calculate
the concurrence and tangle. This is the content of the following Theorem.

\noindent
{\bf Theorem:} \emph{The concurrence $C(\rho _{f})$ resp. tangle $\tau (\rho _{f})$ for the
Werner states $\rho _{f}$ of Eq.~(\ref{wernerstate}) are given by
\begin{equation}
\left\{
\begin{array}{l}
C(\rho _{f})=-f, \\[1mm]
\text{resp.} \\[1mm]
\tau (\rho _{f})=f^{2}.
\end{array}
\right.  \label{resultoftheorem}
\end{equation}
for $f<0$ and $C(\rho _{f})=\tau (\rho _{f})=0$ for $f\geq 0$.}

\vspace*{12pt}
\noindent
{\bf Proof:} It is evident that both
concurrence and tangle will be $0$ according to the convex hull construction of
Eqs.~(\ref{tmin}) and (\ref{cmin}) for $f\geq 0$. We focus on
the case where $-1\leq f< 0$, which implies that the Werner states are
entangled. For any pure state $\sigma $ of Eq.~(\ref{Schmidt}) satisfying $%
P\sigma =\rho _{f}$, we have already found that the minimal admissible
values for $\tau (\sigma )$ and $C(\sigma )$ are given by Eqs.~(\ref{tmin})
and (\ref{cmin}). The optimal choice for $\sigma ={\left\vert \Phi
\right\rangle }{\left\langle \Phi \right\vert }$ is given by Eq.~(\ref%
{solution}).

Now we can compute the convex hull of the function $C(\sigma )$ (or $\tau
(\sigma )=C^{2}(\sigma )$) through the results of
Eqs.~(\ref{measure}), (\ref{cohull}) and (\ref{convexhull})). We have:
\begin{equation}
\left\{
\begin{array}{lll}
\frac{\partial C(\sigma )}{\partial f}=-1<0, & \text{ \ \ \ \ \ \ \ \ } &
\frac{\partial ^{2}C(\sigma )}{\partial f^{2}}=0, \\[2mm]
\frac{\partial C^{2}(\sigma )}{\partial f}=2f<0, & \text{ \ \ \ \ \ \ \ \ } &
\frac{\partial ^{2}C^{2}(\sigma )}{\partial f^{2}}=2>0.
\end{array}
\right.
\end{equation}
Thus both $C(\sigma )$ and $C^{2}(\sigma )$ are monotonically convex
functions of $f$. For the Werner states $\rho _{f}$, which is a convex
combination of the states $\sigma $, one has naturally the results of Eqs.~(%
\ref{resultoftheorem}) according to the convex hull construction. \hfill \rule%
{1ex}{1ex}

It is shown in \cite{Vollbrecht-Werner01} that any pure state in
the optimal decomposition that achieves EOF has the form of
Eq.~(\ref{solution}). The solution can also be rephrased to have
Schmidt rank 2 and Schmidt
coefficients $\mu _{1}=(1+\sqrt{1-f^{2}})/2$ and $\mu _{2}=(1-\sqrt{1-f^{2}}%
)/2$. Thus the optimal decomposition for achieving concurrence and
tangle also achieves EOF at the same time. This shows that all of
the three entanglement measures share a common important feature,
namely to give the same values for {\em every} pure state in the
optimal ensemble decomposition. In addition, the relation shown in
Eq.~(\ref{WernerEOF}) that EOF is a monotonically increasing
function of the concurrence holds naturally, since every pure
state in the optimal ensemble decomposition has the \emph{same}
Schmidt number 2. This is similar to the two qubits case
\cite{Hill-Wootters97,Wootters98} where every pure state in an
optimal ensemble decomposition does have the same value of
concurrence or EOF. Our results thus give the first rigorous proof
for the common ``belief" that, for Werner states, $-f$ plays
exactly the role of concurrence.

In summary, we have given an entirely analytic
derivation of the concurrence and tangle for the Werner states.
Our results show that the concurrence, tangle and entanglement of
formation have the same optimal
decomposition. This is very different from the
isotropic case \cite{Terhal-Voll2000,Rungta-Caves03}, where the
tangle and EOF have a similar behavior while the concurrence behaves
in a completely different manner. This implies that the Werner states have
a more subtle entanglement structure than the isotropic states,
though basically they are partial transpositions of each other,
with respect to one subsystem in some parameter ranges
\cite{Vollbrecht-Werner01}. Since
concurrence is a good entanglement measure and can reveal many
important physical features of the systems involved,
our results would shed new light on a deeper
understanding of entanglement.

\vspace{0.2cm}
\textbf{Acknowledgments.}
K.C. gratefully acknowledges support from the Alexander von Humboldt
Foundation. This work has been supported the Deutsche Forschungsgemeinschaft
SFB611 and German(DFG)-Chinese(NSFC) Exchange Program 446CHV113/231. K.C.
also thanks hospitality of Department of Mathematics in Capital Normal
University where part of this work was finished.


\begin{thebibliography}{99}
\bibitem{nielsen} M.A. Nielsen and I.L. Chuang, Quantum Computation and
Quantum Information, Cambridge University Press, Cambridge, 2000.


\bibitem{Hill-Wootters97} S. Hill and W.K. Wootters, Phys. Rev. Lett.
\textbf{78}, 5022 (1998).

\bibitem{Wootters98} W.K. Wootters, Phys. Rev. Lett. \textbf{80}, 2245
(1998).

\bibitem{Osterloh02-Wu04} A. Osterloh, L. Amico, G. Falci and R. Fazio, Nature \textbf{416},
608 (2002); L.-A. Wu, M.S. Sarandy, and D.A. Lidar, Phys. Rev. Lett. \textbf{%
93}, 250404 (2004).


\bibitem{Ghosh2003} S. Ghosh, T.F. Rosenbaum, G. Aeppli, S.N. Coppersmith,
Nature \textbf{425}, 48 (2003); V. Vedral, Nature \textbf{425}, 28 (2003).

\bibitem{Carvalho05} A.R.R. Carvalho, F. Mintert, and A. Buchleitner,
Phys. Rev. Lett. \textbf{93}, 230501 (2004).

\bibitem{Rungta01-AlbeverioFei01} A. Uhlmann, Phys. Rev. A \textbf{62},
032307 (2000); P. Rungta, V. Buzek, C.M. Caves, M. Hillery, and G.J.
Milburn, Phys. Rev. A \textbf{64}, 042315 (2001); S. Albeverio and S. M.
Fei, J. Opt. B: Quantum Semiclassical Opt. \textbf{3}, 223 (2001).

\bibitem{Audenaert01} K. Audenaert, F. Verstraete, and B.~De Moor, Phys.
Rev. A \textbf{64}, 052304 (2001).

\bibitem{Multi-variables} P. Badziag, P. Deuar, M. Horodecki, P. Horodecki,
and R. Horodecki, J. Mod. Opt. \textbf{49}, 1289 (2002);
H. Fan, K. Matsumoto, and H. Imai, J. Phys. A: Math. Gen.
\textbf{36}, 4151 (2003); G. Gour, Phys. Rev. A \textbf{71}, 012318 (2005).

\bibitem{multigen}
S.J. Akhtarshenas, J. Phys. A: Math. Gen. \textbf{38}, 6777 (2005);
F. Mintert, M. Ku\'s, and A. Buchleitner,
Phys. Rev. Lett. \textbf{95}, 260502 (2005).

\bibitem{lowerbound} P.X. Chen, L.M. Liang, C.Z. Li and M.Q. Huang, Phys. Lett. A \textbf{295}
175 (2002); E. Gerjuoy, Phys. Rev. A \textbf{67} 052308 (2003); A. {\L }ozi{%
\'n}ski, A. Buchleitner, K. Zyczkowski, and T. Wellens, Europhys. Lett. \textbf{62} 168 (2003).


\bibitem{Mintert04-MintertPhD} F. Mintert, M. Ku\'s, and A. Buchleitner,
Phys. Rev. Lett. \textbf{92}, 167902 (2004); F. Mintert, Ph.D. thesis,
Measures and dynamics of entangled states, Munich University, Munich, 2004;
F. Mintert, A.R.R. Carvalho, M. Ku\'s, and A. Buchleitner,
Phys. Rep. \textbf{415}, 207 (2005).



\bibitem{Chen-Albeverio-Fei-PRL200504} K. Chen, S. Albeverio, and S.M. Fei,
Phys. Rev. Lett. \textbf{95}, 040504 (2005).

\bibitem{Rungta-Caves03} P. Rungta and C.M. Caves, Phys. Rev. A \textbf{67},
012307 (2003).

\bibitem{Manne-Caves2005} K.K. Manne and C.M. Caves,
e-print: quant-ph/0506151.

\bibitem{werner89} R.F. Werner, Phys. Rev. A \textbf{40}, 4277 (1989).

\bibitem{Vollbrecht-Werner01} K.G.H. Vollbrecht and R.F. Werner, Phys. Rev.
A \textbf{64}, 062307 (2001).

\bibitem{wernerpreparation} Y.S. Zhang, Y.F. Huang, C.F. Li, and G.C. Guo,
Phys. Rev. A \textbf{66}, 062315 (2002).

\bibitem{Rudolph02} O. Rudolph, quant-ph/0202121.

\bibitem{WoottersQIC2001} W.K. Wootters, Quant. Inf. Comp. \textbf{1}, 27
(2001).

\bibitem{Coffman2000} V. Coffman, J. Kundu, and W.K. Wootters, Phys. Rev. A
\textbf{61}, 052306 (2000).

\bibitem{Osborne2002} T.J. Osborne, Phys. Rev. A \textbf{72}, 022309 (2005).

\bibitem{Terhal-Voll2000} B.M. Terhal and K.G.H. Vollbrecht, Phys. Rev.
Lett. \textbf{85}, 2625 (2000).

\bibitem{Soojoon2003} S. Lee, D.P. Chi, S.D. Oh, and J. Kim, Phys. Rev. A.
\textbf{68}, 062304 (2003).

\bibitem{Virmani-PlenioPLA2000} S. Virmani and M.B. Plenio, Phys. Lett. A
\textbf{268}, 31 (2000).
\end{thebibliography}
\end{document}